\documentclass[twocolumn,superscriptaddress]{revtex4}
\usepackage{graphicx}
\usepackage{times}
\usepackage{textcomp}
\usepackage{amsmath}
\usepackage{amssymb}
\usepackage{units}
\usepackage{stmaryrd} 
\usepackage[compact]{titlesec}
\titlespacing{\section}{0pt}{*0}{*0}
\titlespacing{\subsection}{0pt}{*0}{*0}
\titlespacing{\subsubsection}{0pt}{*0}{*0}

\begin{document}
\preprint{0}

\title{Angle-resolved and core-level photoemission study of interfacing the topological insulator Bi$_{\mathbf{1.5}}$Sb$_{\mathbf{0.5}}$Te$_{\mathbf{1.7}}$Se$_{\mathbf{1.3}}$ with Ag, Nb and Fe
}

\author{N. de Jong}
\email{n.dejong@uva.nl} 
\address{Van der Waals-Zeeman Institute, Institute of Physics (IoP), University of Amsterdam, Science Park 904, 1098 XH, Amsterdam, the Netherlands}

\author{E. Frantzeskakis}
\address{Van der Waals-Zeeman Institute, Institute of Physics (IoP), University of Amsterdam, Science Park 904, 1098 XH, Amsterdam, the Netherlands}

\author{B. Zwartsenberg}
\address{Van der Waals-Zeeman Institute, Institute of Physics (IoP), University of Amsterdam, Science Park 904, 1098 XH, Amsterdam, the Netherlands}

\author{Y. K. Huang}
\address{Van der Waals-Zeeman Institute, Institute of Physics (IoP), University of Amsterdam, Science Park 904, 1098 XH, Amsterdam, the Netherlands}

\author{D. Wu}
\address{Van der Waals-Zeeman Institute, Institute of Physics (IoP), University of Amsterdam, Science Park 904, 1098 XH, Amsterdam, the Netherlands}

\author{P. Hlawenka}
\address{Helmholtz-Zentrum Berlin f\"{u}r Materialien und Energie, Albert-Einstein-Strasse 15, 12489 Berlin, Germany}

\author{J. Sa\'{n}chez-Barriga}
\address{Helmholtz-Zentrum Berlin f\"{u}r Materialien und Energie, Albert-Einstein-Strasse 15, 12489 Berlin, Germany}

\author{A. Varykhalov}
\address{Helmholtz-Zentrum Berlin f\"{u}r Materialien und Energie, Albert-Einstein-Strasse 15, 12489 Berlin, Germany}

\author{E. van Heumen}
\address{Van der Waals-Zeeman Institute, Institute of Physics (IoP), University of Amsterdam, Science Park 904, 1098 XH, Amsterdam, the Netherlands}

\author{M. S. Golden}
\email{m.s.golden@uva.nl} 
\address{Van der Waals-Zeeman Institute, Institute of Physics (IoP), University of Amsterdam, Science Park 904, 1098 XH, Amsterdam, the Netherlands}


\begin{abstract}

Interfaces between a bulk-insulating topological insulator (TI) and metallic adatoms have been studied using high-resolution, angle-resolved and core-level photoemission.
Fe, Nb and Ag were evaporated onto Bi$_{\mathbf{1.5}}$Sb$_{\mathbf{0.5}}$Te$_{\mathbf{1.7}}$Se$_{\mathbf{1.3}}$ (BSTS) surfaces both at room temperature and 38K.
The coverage- and temperature-dependence of the adsorption and interfacial formation process have been investigated, highlighting the effects of the overlayer growth on the occupied electronic structure of the TI.
For all coverages at room temperature and for those equivalent to less than 0.2 monolayer at low temperature all three metals lead to a downward shift of the TI's bands with respect to the Fermi level. At room temperature Ag appears to intercalate efficiently into the van der Waals gap of BSTS, accompanied by low-level substitution {\bf for} the Te/Se atoms of the termination layer of the crystal. This Te/Se substitution with silver increases significantly for low temperature adsorption, and can even dominate the electrostatic environment of the Bi/Sb atoms in the BSTS near-surface region.     
On the other hand, Fe and Nb  evaporants remain close to the termination layer of the crystal. On room temperature deposition, they initially substitute isoelectronically for Bi as a function of coverage, before substituting for Te/Se atoms. 
For low temperature deposition, Fe and Nb are too immobile for substitution processes and show a behaviour consistent with clustering on the surface.
For both Ag and Fe/Nb, these differing adsorption pathways still lead to the qualitatively similar and remarkable behavior for low temperature deposition that the chemical potential first moves upward (n-type dopant behavior) and then downward (p-type behavior) on increasing coverage. \cite{correction} 
\end{abstract}

\maketitle

\section*{Introduction}

Topological insulators (TIs) are a novel material class characterized by topologically-protected electronic states at their interfaces with an ordinary material, such as vacuum. These unique electronic states, known as topological surface states (TSS) exhibit a chiral spin arrangement in which the electron momentum is locked to the spin. They are therefore of high potential in spintronic applications \cite{Hasan2010T,Qi2011T}.  However, in order to use the remarkable properties of the TSS in useful devices, the bulk conductivity has to be small compared to that of the surface. Unfortunately, the most commonly studied TIs, Bi$_2$Se$_3$ and Bi$_2$Te$_3$, are intrinsically doped by Se vacancies and Te-Bi anti-site defects respectively \cite{Cava2013C,Ren2010L,Butch2010S}, both of which induce high bulk conductivity.\\
\indent
Controlled changes of the bulk stoichiometry can result in an effective reduction of the bulk conductivity  \cite{Checkelsky2009Q,Hsieh2009A,Analytis2010T,Ren2011O}. A comprehensive study of the material Bi$_{2-x}$Sb$_{x}$Te$_{3-y}$Se$_y$ has revealed that  Bi$_{1.5}$Sb$_{0.5}$Te$_{1.7}$Se$_{1.3}$  (BSTS) is a good bulk insulator with resistivities at low temperature as high as 4 $\Omega$ cm \cite{Ren2011Op} or 10 $\Omega$ cm \cite{Pan2014L} . BSTS belongs to the tetradymite group of materials and thus has the same rhombohedral crystal structure as Bi$_2$Se$_3$, with three quintuple layers in the unit cell. For BSTS the layer sequence within the quintuple layers is expected to be: Te/Se - Bi/Sb - Se - Bi/Sb - Te/Se \cite{Nakajima1963J,Cava2013C} meaning that the surface termination is of mixed Te and Se character. Each Bi/Sb atom forms  $\sigma$-bonds with the surrounding Te/Se atoms, thus in effect sitting in the center of an almost regular coordination octahedron \cite{Krebs1968,Vasko1974A,Cava2013C}.\\
\indent
An important step towards future topological devices involves a comprehensive understanding of the microscopic phenomena that take place when TIs and different materials form stable interfaces. This is of course important because in a device one has to make a connection between the TI and other electronics. Although such phenomena have been studied for a range of different adsorbate materials on Bi$_2$Se$_3$ and Bi$_2$Te$_3$ \cite{Wray2010A,Scholz2012T,Zhu2011R,Valla2012P,Honolka2012I,West2012I,Vobornik2014O}, relevant data are missing on bulk insulating 3D-TI compounds, such as BSTS. Here, results from angle resolved photoelectron spectroscopy (ARPES) and core level spectroscopy are combined to allow a better understanding of the interface and of the evaporated adsorbate-induced effects on the topological band structure.
\\
\indent
We investigate the interfaces of BSTS with silver, niobium and iron. These materials represent three categories of interest: 1) conventional metals (Ag) as a typical contact material for electrical connections in TI devices, 2) magnetic metals (Fe) that may be used as a ferromagnetic metal usable to contact magnetically-doped TI's and 3) superconducting materials (Nb) which might be used in fabricating TI-superconducting junctions \cite{Veldhorst2012E,Veldhorst2012J}. We focus also on the difference between the behavior of these adsorbates at room temperature and at low temperature (38K). 
\\

\begin{figure*}
  \centering
  \includegraphics[width = 17.6 cm]{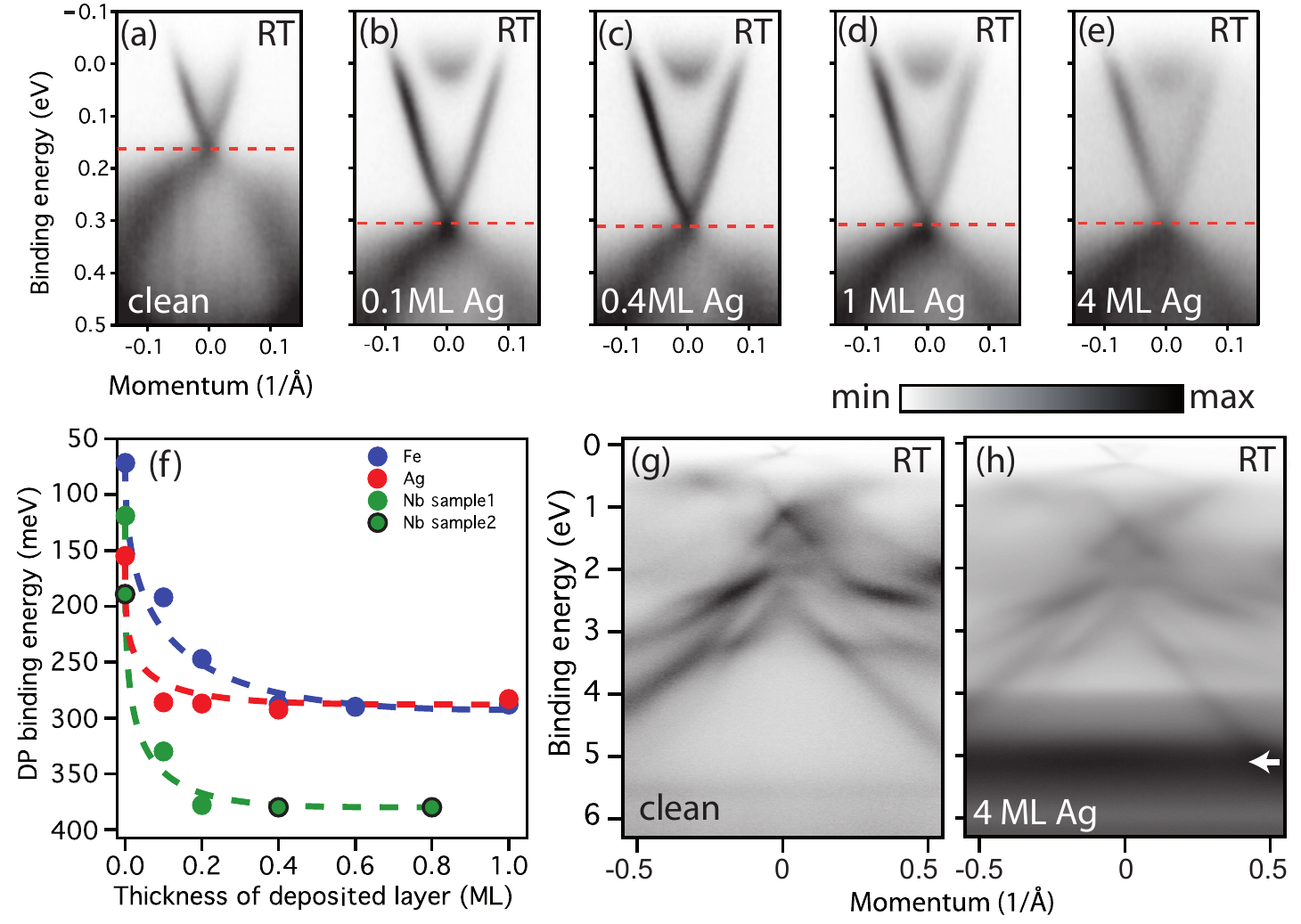}
 \caption{{\bf Electronic structure of Bi$_{1.5}$Sb$_{0.5}$Te$_{1.7}$Se$_{1.3}$ upon adsorption of silver at room temperature, and comparison with Nb and Fe deposition.} I($E,k$) images for the region near E$_F$  as a function of the amount of evaporated Ag: (a) Clean surface, (b) 0.1 ML, (c) 0.4 ML (d) 1.0 ML, (e) 4.0 ML, where ML stands for monolayer (f) Binding energy of the Dirac point (DP) as function of Ag, Nb and Fe thickness, dashed lines are guides to the eye. Error bars on the energy of the DP depend on the coverage, ranging from $\pm$ 5 meV for the clean surface to $\pm$ 15 meV for 4 ML of Ag (the symbol size in panel [f] corresponds to 17 meV). (g,h) Valence band I$(k,E)$ images for (g) the clean surface and (h) after evaporation of 4 ML of Ag. Samples were kept at room temperature during evaporation and measurements, the latter being performed using a photon energy of 27 eV. }
\label{Fig1}
\end{figure*}

\section*{Experimental}

High quality BSTS single crystals were grown in Amsterdam using the Bridgman technique. High purity elements were melted in evacuated, sealed quartz tubes at 850\textdegree C and allowed to mix for 24 hours before cooling. The cooling rate was 3\textdegree C per hour. Samples were cleaved in UHV (pressure \textless $5\times10^{-10}$ mbar) at either 38K or room temperature (RT).\\
\indent
High-purity (99.995\%) Nb and Fe were evaporated using a commercial e-beam evaporator (EFM3 Focus GmbH) which was calibrated using its internal flux monitor and a quartz microbalance. Evaporation rates of order 0.1 monolayer (ML) per minute were used. Ag was evaporated using a tungsten filament, which was also calibrated using the same quartz microbalance. Evaporation of all materials was with the sample held at either room temperature or at 38 K. \\
\indent
ARPES and core level photoemission measurements were performed at the UE112-PGM-2a-1\string^2 endstation of the BESSY II synchrotron radiation facility at HZB using a Scienta R8000 hemispherical electron analyzer and a six-axis manipulator. All ARPES measurements presented here were measured using a photon energy of 27 eV. The total energy resolution was 30 meV, and the angular resolution was 0.2$^{\circ}$, resulting in 0.0085 \AA$^{-1}$ resolution in momentum space. For the core level photoemission data a photon energy of 130 eV was used, yielding an overall energy resolution of 50 meV. The pressure during the measurements was $1.0\times10^{-10}$ mbar and they were performed at room temperature and at 38K.\\

\section*{Results and Discussion}
\subsection*{I. ARPES data}
\subsubsection*{Room temperature deposition and measurement}
Fig. 1 presents a summary of the ARPES results after the deposition and measurement of Ag, Fe and Nb on Bi$_{1.5}$Sb$_{0.5}$Te$_{1.7}$Se$_{1.3}$ at room temperature. The near-E$_{F}$ electronic structure of BSTS as a function of increasing Ag layer thickness is shown in panels (a) to (e). The observed bands shift $\sim$130 meV to higher binding energy for the smallest coverage of 0.1 ML. For higher coverages no further energy shift is observed. Fig. 1f shows the energy of the Dirac point as a function of the amount of deposited material at RT for all three metals used in this study (the ARPES spectra after RT deposition with Fe and Nb are shown in Appendix A). Comparing the three metals, a similar behavior is observed: for low coverage the bands shift strongly (between 130 and 250 meV) to higher binding energy \footnote{The differing initial values of the Dirac point binding energy visible in Fig. 1f and 2f are due to the different samples/cleaves involved being measured at slightly different times over which the adsorbate-induced downward band bending in BSTS has been operative \cite{Frantzeskakis2015D}}. For higher coverage this shift saturates and the binding energy of the near E$_F$ features remains stable. The deposition thickness and the binding energy at which this saturation takes place is different for each adatom species.\\
\indent
 A similar energy shift is often observed in Bi$_2$Se$_3$ after exposure to residual gas atoms in UHV \cite{Hsieh2009A,Park2010Q,King2011L,Benia2011R,Bianchi2011S} or when the surface is deliberately decorated with alkali \cite{Zhu2011R,King2011L,Bianchi2012R} or other metals \cite{Wray2010A,Scholz2012T,Wray2011E,Valla2012P}. Deliberate or indirect decoration with adatoms is at the origin of downward band bending which -in the case of Bi$_2$Se$_3$- can be strong enough to create a potential well at the surface confining the conduction and valence band states \cite{King2011L,Bianchi2012R}. Such a time-dependent shift of the bands at the surface to higher binding energy has also been observed for BSTS \cite{Frantzeskakis2015D}. For the experiments reported here the first deposition was done long before the saturation in the shift caused by residual gas in the UHV chamber. After the surface has been deliberately decorated with adatoms, no more shifting due to the adsorption of residual gas atoms is observed.\\
 \indent 
Our room temperature ARPES results with low coverage (0.1 ML) on BSTS shown in both Fig. 1 and Appendix A can be explained using a similar reasoning. Ag, Nb and Fe atoms which adsorb on the surface act as electron donors, leaving a positively charged ion near the surface and the resulting downward band bending shifts the near-surface bands to higher binding energies. A parabolic state appears inside the Dirac cone after evaporating 0.1 ML of Ag (Fig. 1b). This is ascribed to the band-bent conduction band which is prone to quantization on increasing the potential gradient at the surface of BSTS, similar to what has been observed in Bi$_2$Se$_3$ \cite{King2011L,Bianchi2012R}. In comparison to the Bi$_2$Se$_3$ case \cite{King2011L}, downward band bending at the surface of BSTS saturates at lower binding energies and no Rashba-Bychkov splitting has been observed for the states emerging below E$_{F}$. This suggests a weaker surface potential gradient for BSTS in comparison to Bi$_2$Se$_3$.\\
\indent
Upon silver adsorption, changes can also be seen in the valence band region shown in Figs. 1g and 1h: the 4d states of Ag  become visible as a non-dispersing feature just below 5 eV binding energy (indicated in Fig. 1h by the small black arrow). As one would expect, the intensity of the Ag 4d feature is observed to increase as a function of the coverage. \\
\indent
There are three main types of sites that these metals can occupy after deposition so as to act as electron donors:
\begin{enumerate}
\item As adatoms on top of the surface. 
\item Intercalated between quintuple layers in the van der Waals gap.
\item At interstitial sites within a quintuple layer
\end{enumerate}
The preferred position of the metal adatoms will be highly dependent on which atom is used. For example, for Bi$_2$Se$_3$ it was shown that Ag adatoms deposited at room temperature possess enough mobility to move to step edges and intercalate into the van der Waals gap \cite{Otrokova2012E,Gosalves2014L,Ye2011R}, as is also the case for Cu adsorption \cite{Wang2011S}. In contrast, Fe was observed to prefer to occupy interstitial sites or substitute for Bi within the quintuple layers of Bi$_2$Se$_3$ \cite{Schlenk2013C} or Bi$_2$Te$_3$ \cite{West2012I}. As will be discussed later in this section, these substitutional defects do not act as electron donors \cite{Schlenk2013C}. \\
\indent
Here we would like to point out the observation that the maximum nominal layer thickness deposited at room temperature that still allows observation of a Dirac cone with ARPES involving VUV photons, is much higher for Ag (4 ML, Fig. 1e) than it is for Fe and Nb ($\sim$1 ML, Figs. A1a and A1c).
While this difference could reflect a lower sticking coefficient for silver at room temperature, in the light of the previous studies on Ag and Cu adsorption on Bi$_2$Se$_3$ mentioned above, the more likely scenario is one in which the Ag deposited on BSTS at room temperature moves to step edges and intercalates into the van der Waals gap. This makes the silver a much less effective attenuator of the ARPES signal from the TI electronic states.
As the transition metals are generally considered to be more reactive than silver, and interstitial and substitutional sites are energetically preferable for Fe (and by extension and Nb), these species remain closer to the surface, thereby interfering more with the ARPES signal from the underlying BSTS.
A further argument against significant occupation of Bi(Sb) sites by silver is that the latter would involve a strong p-type doping, as Ag brings one 4s electron to the problem, compared to the three 6p(5p) electrons for Bi(Sb)  \cite{Vasko1974A}. Thus, a Ag$_{\text{Bi,Sb}}$ scenario does not agree with the shift of the ARPES features to high binding energy as silver is deposited. Thus, at 0.1ML, enough of the silver ions remain as adatoms at the BSTS surface so as to result in the downward shift of the near-surface band structure observed in Fig. 1b. Further silver disposition at room temperature leaves the band bending unaltered, and, as discussed above, the silver most probably acts as an intercalant in the van der Waals gap.

\begin{figure*}
\centering 
  \includegraphics[width = 18 cm]{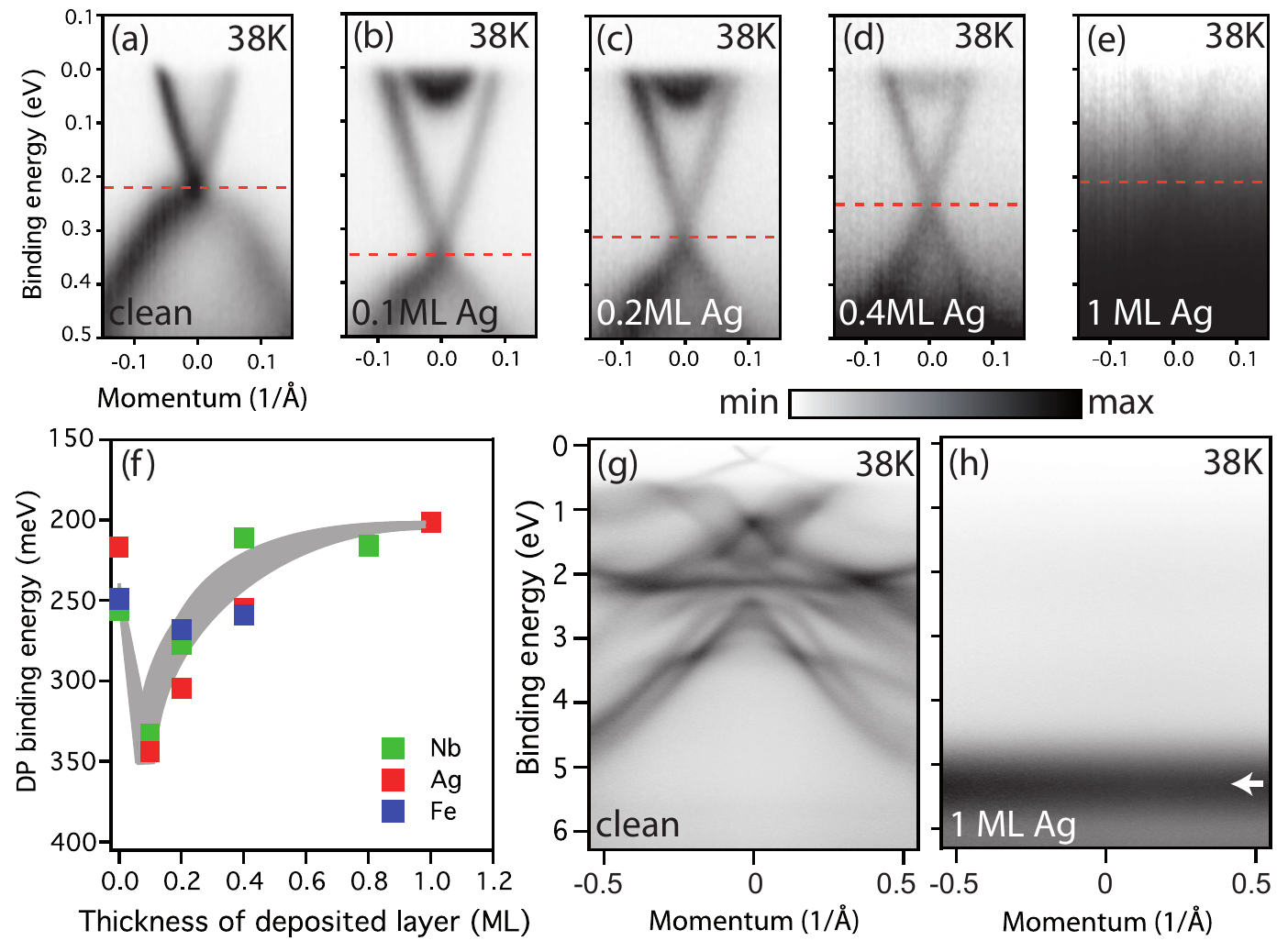}
 \caption{{\bf Electronic structure of Bi$_{1.5}$Sb$_{0.5}$Te$_{1.7}$Se$_{1.3}$ upon adsorption of silver at 38K, and comparison with Nb and Fe deposition.} (a-e): Near-E$_F$ dispersion relations for: (a) the clean BSTS surface, (b) 0.1 ML Ag, (c) 0.2 ML Ag (d) 0.4 ML Ag, (e) 1.0 ML Ag, (ML stands for monolayer). (f) Binding energy of the Dirac point (DP) as function of Ag, Nb and Fe evaporation thickness, with the shaded region acting as a guide to the eye.  Error bars on the energy of the DP depend on the coverage, ranging from $\pm$ 5 meV for the clean surface to $\pm$ 20 meV for 1 ML of Ag (the symbol size in panel [f] corresponds to 13 meV). (g,h) Valence band $I(k,E)$ images for (g) the clean surface and (h) after evaporation of 1 ML of Ag, for which the arrow indicates the emission from Ag4d states. For all data shown the samples were kept at 38K during evaporation and measurements, the latter being performed with a photon energy of 27 eV. 
}
\label{Fig2}
\end{figure*}

In Bi$_2$Se$_3$, Fe substitution at the Bi site has been shown to be charge neutral \cite{Schlenk2013C}. Calculations and experiments for Fe adsorption on both Bi$_2$Te$_3$ \cite{West2012I} and Bi$_2$Se$_3$ \cite{Schlenk2013C} suggest that at room temperature,  Fe$_{\text{Bi}}$ is the most energetically favorable outcome. Thus, after an initial downward band bending at 0.1ML coverage, the lack of a significant chemical potential shift for nominal coverages up to 1ML would be consistent with the formation of charge-neutral Fe$_{\text{Bi,Sb}}$ in BSTS.\\
\indent
Turning now to Nb deposition on BSTS at RT, from the point of view of the ARPES data, Nb is observed to show similar behavior as Fe (Fig. 1f, A1c and A1d). It induces somewhat stronger downward band bending than Fe when adsorbed at the 0.1 ML level, but then yields a steady chemical potential over the studies coverage range (0.8ML). As for iron, niobium is not expected to intercalate significantly into the van der Waals gap but rather occupy sites close to the surface (adsorbed on top of the surface, built into the surface termination layer or at Bi substitutional sites). \\ \indent
What is the take-home message from our ARPES data for those interested in making TI devices from BSTS?
For the technically-relevant room temperature deposition of up to 1ML Ag, Fe and Nb on BSTS, all three metals lead to a downward shift of the BSTS bands in the near-surface region.
For silver and iron this leads to a Dirac point energy of $\sim$275 meV below E$_F$, and for niobium, the spin compensation point is located some $\sim$375 meV below E$_F$. For all three metals, the bulk conduction band states are partially occupied in the near-surface region, meaning besides the TSS, there is a second, topologically trivial conduction channel.\\
\indent
 
\subsubsection*{Deposition and measurement at 38K (ARPES)}
Fig. 2 shows the results when the metal deposition (and ARPES measurement) is performed at 38K. Similar to the room temperature case, the initial shift of the chemical potential is upwards for small coverages, as can be seen in Fig. 2b, signifying donor-like behavior.
However, for greater coverages the situation is drastically different to that of room temperature in that in the low T data for all three metals, the bands shift back to lower binding energies (Figs. 2c to 2f). This marks a transition in the doping behavior of the deposited metal species: from donor- to neutral or acceptor-like. \\
\indent  
Comparison of the panels (e) and (g-h) of Figs. 1 and 2 shows that the low temperature data present a much higher background signal compared to the TI-related emission for higher coverages. At 38K, the Ag 4d states dominate the valence band spectrum completely, already at 1 ML. This difference we observe between the behaviour for Ag on BSTS between room and low temperature is similar to that reported for Ag \cite{Otrokova2012E,Gosalves2014L,Ye2011R} and Cu \cite{Wang2011S} on Bi$_2$Se$_3$.
For Fe and Nb, increase in the background signal at higher coverages for 38 K deposition compared to  room temperature is also observed (see Appendix A), although these effects are less marked than for Ag.
In general, our results for BSTS for deposition at 38K echo the expectations from binary TI's that almost all deposited adatoms will be 'frozen' in place on top of or in the surface layer, as they do not possess the mobility to move into subsurface sites \cite{Otrokova2012E,Gosalves2014L,Schlenk2013C,West2012I,Wang2011S,Honolka2012I}.\\
\indent
Returning to the low temperature reversal of the doping character of all three metals as function of deposition thickness. 
For the Nb and Ag case, the 0.1 ML evaporation at 38K pins the deep minimum in the `tick-shape' variation (grey guide to the eye in Fig. 2f) of the Dirac energy vs. coverage. For Fe, we do not have data for 0.1 ML at 38K, but from the other ARPES data and the core level data that will be shown shortly, Fe can be seen to behave similarly to Nb, so it could be assumed that for 0.1 ML Fe, the Dirac point energy could also be of order 340 meV. Concentrating on the data points we do have at our disposal, it can be seen that the Fe data do also show a reversal of the doping character between coverages of 0 and 0.2 ML vs. between 0.2 and 0.4 ML. Without the 0.1 ML data point for Fe, the question of whether this reversal is less pronounced in the case of iron deposition is something of a moot point.
\\
We note that although the formation of substitutional defects could explain this acceptor-like behavior, in the binary systems substitutional defects are generally strongly suppressed at low temperature \cite{Otrokova2012E,Gosalves2014L,Schlenk2013C}. Of possible relevance here is the fact that reversal of the effect on the surface band bending and thus apparent doping behavior (from n- to p-type) for gold on titanium dioxide surfaces has been observed \cite{Zhang2011E} on going from single adatoms on the surface to the formation of clusters. By inference, these kind of effects could also play a role in the case of the three elements deposited here at low temperature on BSTS.\\
In the literature, the deposition of Fe on Bi$_2$Se$_3$ at low temperature has yielded apparently contradictory results as regards the doping behavior: with acceptor-like behavior \cite{Scholz2012T} for 0.2 and 0.4 ML and donor-like for 1\% of a ML \cite{Honolka2012I}. The results from BSTS shown in Fig. 2 suggest that this difference in the data from the selenide could be caused by the reversal of the doping character as a function of the amount of Fe deposited on the surface.\\
\indent
Finally, comparing the RT and 38K deposition+ARPES experiments one can also conclude that at low temperatures by choosing an appropriate coverage the energy of the topological surface states can be tuned such that the conduction band states of BSTS can be pushed back above E$_F$ (Fig. 2e). This is of fundamental interest as the conduction band of BSTS becomes occupied with electrons at the surface simply following a few hours of exposure to UHV \cite{Frantzeskakis2014M,Frantzeskakis2015D}.\\ \indent

\subsection*{II. Core level photoemission data}
\subsubsection*{Room temperature deposition and measurement}
In the following, we use core-level spectroscopy to provide information on changes in the bonding and electrostatic environment of deposited Ag, Nb or Fe atoms.
Figs. 3a and 3b show a compilation of the data for the Bi5d and Sb4d core-levels (both atoms share the same planes in the BSTS crystal structure) for Fe and Nb, respectively, for deposition and measurement at both room temperature and 38K. Fig. 3c shows the analogous data for silver deposition at RT. 
In Figs. 3a-3c, a low-binding-energy shoulder is seen for the Bi5d and Sb4d core-levels, becoming clearly visible for a coverage of 2ML or higher. These low-binding-energy features, split off from the main lines by $\sim$0.9 eV indicate a more electropositive environment for some of the Bi(Sb) atoms in the surface region of the BSTS crystal, presumably for those close to where the Ag, Nb or Fe atoms have been incorporated into the crystal.
A zoom of the Bi5d lines for the three RT-deposited metals is shown in Fig. 3e.\\
\indent 
For low coverages (well below 1 ML), the BSTS core-level data show no additional features associated with significant changes in the electrostatics within the near-surface region. This is in keeping with the situation for Fe on Bi$_2$Se$_3$ and Bi$_2$Te$_3$, in which electrically neutral substitution at the Bi site is reported to occur.\\
\indent
Although for higher coverages the above-mentioned low binding energy shoulder is seen on the Bi/Sb core-level spectra, the Te(Se) 4d(3d) core-level data shown in panel Fig. 3f show no change in oxidation state of the Te or Se at room temperature after deposition of any of the metals used. This point will be returned to in the discussion of the LT data.
Finally, for completeness, we note that the additional intensity at the foot of the Se 3d core-level marked with an arrow for the case of iron (Fig. 3f, 3g, blue arrow in panel [f]) is due to the signals from the Fe3p core level line \cite{Scholz2012T}.

\begin{figure*}
\centering 
  \includegraphics[width = 18 cm]{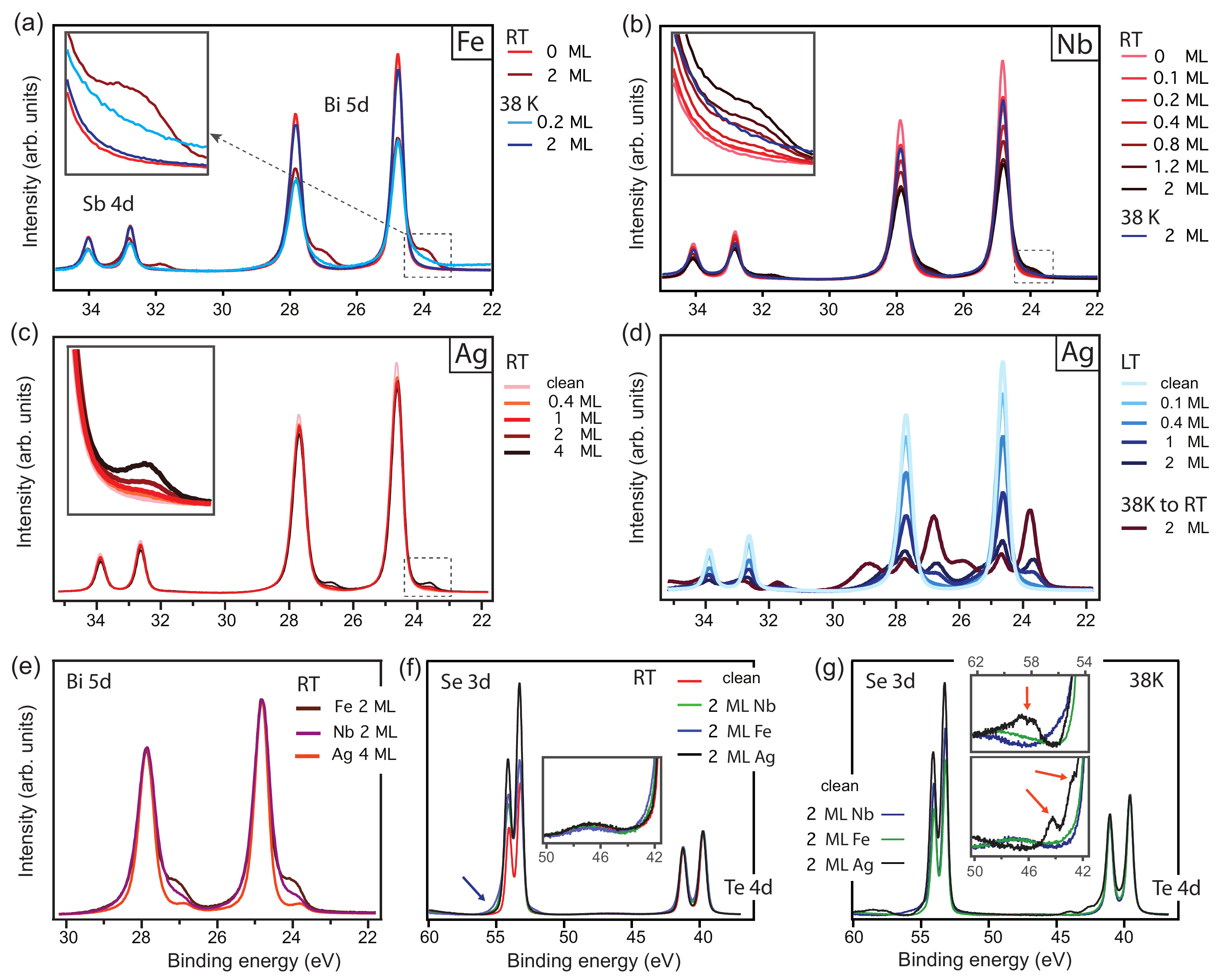}
 \caption{{\bf High-resolution core-level spectra.} (a-b) Bi 5d and Sb 4d core-levels for BSTS decorated with the indicated amounts of (a) Fe, (b) Nb evaporated and measured at room temperature (red curves) or at 38 K (blue curves). Panel (c) shows analogous data for Ag (RT only). The insets in (a-c) show zooms of the indicated regions. (d) Bi 5d and Sb 4d core-levels for BSTS decorated with the indicated amounts of Ag evaporated and measured at 38 K. The bordeaux red trace was recorded from a BSTS sample decorated with 2 ML of Ag at 38 K that was subsequently allowed to heat up to RT before measurement. (e) zoom of the Bi 5d core level lines comparing room temperature deposition (and measurement) for 2 ML of Fe and Nb and 4 ML of Ag. (f) Te 4d and Se 3d core-levels for BSTS decorated with 2 ML of either Fe, Nb and Ag at room temperature. The blue (dark grey) arrow indicates where the Fe 3d emission is situated. (g) Te 4d and Se 3d core-levels for BSTS decorated with 2 ML of either Fe, Nb and Ag at 38K. The insets to panels (f) and (g) show zooms of higher binding-energy regions of the Te 4d and Se 3d features. In the inset to (g), the orange (light grey) arrows indicate signs of oxidation during Ag deposition.
To ease comparison, the spectra have been shifted in energy such that in panels (a-e) the main feature of the Bi 5d levels are aligned. The energy shifts used closely track the chemical potential shifts seen in the ARPES data of Figs. 1, 2 and A1. For panels (a-d) the core-level datasets are normalised across the binding energy range of 0-110eV. For panel (e) the main Bi 5d feature is used for normalisation. In panels (f) and (g), the traces are shifted and normalised to coincide in both energy and intensity for the main Te4d line. All spectra were measured with h$\nu$ = 130 eV.
}
\label{Fig3}
\end{figure*}

\subsubsection*{Deposition and measurement at 38K (core-levels)}
Turning now to deposition and measurement carried out at 38K, the core-level data of Fig. 3a and 3b tell us that for Nb and Fe, even when the deposited overlayer is so thick that the topological surface states are no longer visible in ARPES (\textgreater 1ML for 38K), the Bi5d and Sb4d core level lines show no low-binding-energy shoulders.
For Nb and Fe deposition at 38K, the data of Fig. 3g (and insets) also show a lack of doublets at the high binding energy side of either the Se3d or Te4d core levels.\\
\indent
The data for low-T silver deposition shown in Fig. 3d and 3g are quite different. Firstly, for the Bi/Sb core levels, the 38K Ag case does show low-binding-energy shoulders, and comparison of panels (c) and (d) of Fig. 3 show that the low-BE shoulders for silver at 38K can be much more pronounced than in the silver data for room temperature deposition and measurement. 
For 2 ML of Ag deposited and measured at 38K, there are signs of high-BE features at all four core levels of the BSTS (Bi and Sb in Fig. 3d; Se and Te in Fig. 3g).
This issue will be returned during the discussion of the data of the 38K Ag deposition which was subsequently warmed slowly to room temperature and then measured again.\\
\indent
In the literature, Fe deposition on Bi$_2$Se$_3$ \cite{Scholz2012T} has been shown to lead to low-binding energy features like those seen here in the Bi5d lines (also without changes in the Se3d spectra), but in contrast to BSTS, these new Bi5d features are present for deposition and measurement {\it both} at 8K and RT.
For Bi$_2$Te$_3$ \cite{Vobornik2014O}, not only are low-binding energy features seen in the Bi5d lines upon room temperature deposition of Fe, but new and intense Te4d signals at $\sim$0.5 eV higher binding energy are also observed, attributed to the likely formation of iron telluride (FeTe$_2$) at the surface \cite{Vobornik2014O}.\\
\indent
The surface of Bi$_{\mathbf{1.5}}$Sb$_{\mathbf{0.5}}$Te$_{\mathbf{1.7}}$Se$_{\mathbf{1.3}}$ would, on average, present a termination of 85$\%$ Te and 15$\%$ Se \cite{Nakajima1963J,Cava2013C}.
The fact that no new Te core level signatures are seen in our data indicates that the substitutional behaviour of the evaporants on BSTS resembles more that of Fe on Bi$_2$Se$_3$ \cite{Scholz2012T} than on Bi$_2$Te$_3$ \cite{Vobornik2014O}. Therefore, in line with Ref. \cite{Scholz2012T}, we interpret the low-binding-energy shoulders observed on the Bi and Sb core level lines at room temperature to be mainly the result of the substitution - by Ag, Fe and Nb - of the Se atoms of BSTS. As no new core-level signatures of the displaced Se atoms are visible, they could have evaporated into the UHV environment. If surface tellurium atoms were to be displaced out of the BSTS on substitution, the lower vapour pressure of this element could lead to the formation of tellurium compounds as clusters or islands at the surface, as reported based on high binding energy Te core-level features seen for Fe on Bi$_2$Te$_3$ in Ref. \cite{Vobornik2014O}. From fitting the core level peaks and their respective shoulders, we have compared the area of the shoulder with that of its associated main peak, thereby enabling a rough estimate of the percentage of Bi atoms that have a broken bonds to the Te/Se surface layer. Such an analysis yields a percentage of Bi atoms `missing' a bond to a Te/Se atom to be a little over 2\% for Ag, just under 4\% for Nb, and 12\% for Fe adsorption. Since each Te/Se surface atom has a bond with 3 Bi atoms in the layer below it, these values need to be divided by three to get the percentage of Te/Se atoms which are actually replaced at the surface, which brings these numbers for all three evaporants to well below the nominal Se concentration of 15\% at the outermost surface layer, and is thus consistent with a picture in which mainly the Se is replaced by the deposited adatoms. The reasoning just given above is not applicable to the deposition of Ag at 38K, for which a large shoulder is observed. For this case we estimate 49\% of the Bi atoms below the surface to have broken bonds, meaning 16.33\% of the Te/Se surface atoms are replaced by Ag. Again we come back to this issue when discussing the data of the 38K Ag deposition which was subsequently warmed slowly to room temperature and then measured again.\\
\indent
The BSTS termination surface can be considered to be a linear combination of those of Bi$_2$Te$_3$ and Bi$_2$Se$_3$, and the core-level spectra of metal-adsorbed BSTS do indeed share some common features with those of both Bi$_2$Se$_3$ and Bi$_2$Te$_3$, such as the presence of low-binding-energy features for the Bi5d and Sb4d levels.
However, it is clear from our core-level analysis that as regards the behaviour upon deposition of Fe, BSTS is different to both the pure selenide or telluride. Unlike the latter, no signs of iron telluride (or other chemically-shifted Te [or Se] species) are seen, but unlike the former, deposition of iron - and similar in its behavior - niobium at low temperature leave all the core levels of BSTS essentially unchanged.   \\
\indent
Having described the general characteristics of RT and LT decoration of BSTS with metals, we now turn to discussing the particular behavior of Ag, which is different with respect to that of Nb and Fe as evaporants. 
Considering the valence electron counts of the constituent atoms, replacement of a Te or Se atom with a silver atom would withdraw four electrons housed in the 5p(4p) orbitals of a Te(Se) atom, for them to be replaced by one electron from the Ag4s level, thereby effectively doping the system with three holes.
Formation of a Ag$_{\text{Te,Se}}$ substitutional defect would also break the $\sigma$-bonds that connected the Te(Se) atom to its Bi/Sb coordination octahedron \cite{Krebs1968}.
The Te/Se-Bi/Sb bonds are polar covalent, with greater electron density on the more electronegative Te/Se atoms.
When insertion of a silver atom at the Te/Se site removes such a Te/Se-Bi/Sb bond, the Bi/Sb atoms are robbed of one relatively electronegative bonding partner, getting an electropositive metal in its place. This can be expected to lower the ionisation energy for removal of a core electron at the Bi or Sb sites that abut an incorporated silver atom, giving rise to the observed low-binding-energy shoulders.\\ \indent
For Fe and Nb deposition, these shoulders are only seen for room temperature deposition and measurement. These transition metal atoms are essentially immobile on the surface at low temperatures and are not  able to substitute for Te/Se atoms in the (near) surface region of BSTS at 38K. 
The presence of sizeable low-binding-energy features in the Bi/Sb core level data of Fig. 3d show that the silver located at the surface is able to substitute for Te/Se, even at low temperatures. 
Therefore, although the increased attenuation of the valence band ARPES data of BSTS seen on going from Figs. 1h to 2h, as well as the growth of the Ag4d-related valence band states show that the silver is less nimble at 38K, the fact that it can substitute for what are probably Se atoms in the termination layer shows it still to be mobile at low temperatures.  \\
\indent
These observations suggest that for 1ML and greater nominal coverages there could be two factors contributing to substitution for atoms in the Se/Te termination layer: (i) high effective surface concentration of the adatoms, and (ii) sufficient mobility of the adatoms. 
If one of these two factors goes unfulfilled, significant substitution does not take place.
Deposition at low temperature satisfies the first requirement, whereas deposition at RT satisfies the second requirement.
For Ag, the smallest Bi/Sb low-binding-energy shoulders are seen for room temperature deposition (Fig. 3c), in keeping with the rapid intercalation of the silver away from the termination surface.   
Low T deposition of silver generates good coverage of the surface and even at 38 K the silver is mobile enough to undergo substitution at the Te/Se sites so as to give the largest low-binding-energy component to the Bi/Sb core-level lines (Fig. 3d) of all the data for deposition and measurement at the same temperature.
In contrast to the situation for silver, only RT deposition of the transition metals yields substitution. Their highly effective immobilisation on the surface at low temperature simply preventing substitution.\\
\indent 
To achieve the highest level of metal substitution for Te/Se, both of the factors discussed above should be operative at the same time. 
We can test this hypothesis using a Ag layer generated on BSTS at 38K, with subsequent heating up to RT.
The low-T deposition will give an initially high effective surface concentration of silver (due to the suppression of the intercalation process), followed by increasing mobility of the Ag as the slow sample warm-up takes place (in this case the warm-up was carried out by terminating the cooling of the cryostat, without additional heating).
Repetition of the core level measurements after such a process (with a Ag thickness of 2 ML) yields the bordeaux red trace in Fig. 3c, showing that the low-binding-energy shoulders of the Bi and Sb core levels are even the dominant contribution to the core-level spectrum.
This could suggest that the majority of the Bi and Sb atoms within the $\sim 0.3$ nm probing depth of the core-level photoemission measurements possess missing $\sigma$-bonds to the Te/Se atoms, as the latter have been replaced by silver atoms.\\
\indent
However, a caveat relevant for this last conclusion are the strong, additional features observed at the higher binding energy side of the Bi5d and Sb4d core-levels shown with the dark red curve in Fig. 3d for Ag at 38K. As we have also observed these features on a BSTS sample which was exposed to air (data not shown), we ascribe them to the oxidation of the sample surface, due to the adsorbed gases liberated on the warm-up of the cryostat. This means that the data for the 38K deposition of 2 ML of silver, followed by a slow-warm-up indicate signs of damage/decomposition of the BSTS surface.
Armed with this knowledge, we re-examine the 2ML data for all three metals deposited and measured at 38K. 
On doing so, we pick up no signs of oxidation (nor of substitution) for Fe or Nb. 
However, for the silver case, faint signs of the beginnings of oxidation can be seen as high BE features on all four BSTS core-levels, with the insets of Fig. 3g illustrating clearly that these are only seen (here for the Te4d and Se3d lines) for Ag and not for Fe and Nb.\\
\indent
We finish this section with remarks that combine our observations using ARPES and core-level spectroscopy. 
For room temperature, and coverages $>$0.4 ML, the substitution of evaporant atoms at Se(Te) sites seems to have little or no effect on the chemical potential in the near-surface region.  This can be concluded from comparing the evolution of the near-E$_F$ band structure at high coverages ($ \geq 0.4$ ML) for RT shown in Fig. 1 for silver and Fig. A1 for iron and niobium with the evolution of the low-binding-energy core-level shoulders shown in Fig. 3. On the face of it, this result is surprising, as from the point of view of valence electron counting, Ag substitution for Te/Se  - for example - could be expected to generate holes, shifting the chemical potential down. 
Two possible explanations are the following.
Firstly, the holes generated by Ag$_{\text{Te,Se}}$ defects could be swept to the bulk (out of range of the photoemission measurements) by the band bending potential in the space charge layer.
Secondly, if the excess positive charge were to remain localised at or close to the surface, this could possibly strengthen the downward band bending of the electron bands, with no net shift of the bands with respect to the Fermi level being the result. \\
\indent
For deposition and measurement at 38K, the lowest coverage of 0.1 ML yields a strong upward shift of the chemical potential. This we attribute to downward band bending at the surface due to the evaporate adatoms acting like donors, on top of the termination layer.
On increasing the coverage to 0.2-0.4 ML, the core levels still show no signs of substitution by the evaporants, yet the ARPES data reveal a reversal to p-type behavior, which we attribute to clustering of the Fe, Nb or Ag adatoms \cite{Zhang2011E}.
For Fe and Nb deposited and measured at 38K, further increase in the coverage does not qualitatively change the situation, as their lack of mobility precludes substitution.
For silver coverages $>$0.4 ML, the ARPES shows the bands to shift further towards E$_F$, albeit more slowly than between 0.1 and 0.4 ML coverage.
This could still be the result of clustering, but the core-level data also signal on the one hand, the formation of Ag$_{\text{Te,Se}}$ (expected to generate p-type doping), and, on the other hand, first signs of mild oxidation (also known to p-type dope Bi$_2$Se$_3$ \cite{Chen2010M}).
We remark here that the e-beam evaporation of Fe and Nb led to no oxygen contamination, unlike the resistively-heated-basket used for the silver evaporation (which for 2 ML nominal thickness took 20 minutes).

\section*{Summary and conclusions}
From the data presented, it is clear that interfacing BSTS with these different metals can have large impact on the topological surface bands of the TI. How these surface bands are influenced depends on the material that is used, on the temperature at which the deposition is done and the temperature at which the sample is kept after deposition.\\
\indent
From a joint analysis of ARPES and core level data, we show that although there are major differences between the structure/location of the different adatoms deposited, the effects on the electronic structure of the TI that result are similar. For metals deposited at room temperature the Dirac cone shifts to higher binding energies, with partial population of the bulk the conduction band. This is associated with an increased downward band bending at the surface, caused by n-type doping of the different metals deposited at room temperature. The attenuation of the ARPES signal (and the BSTS core levels) as function of deposition thickness, shows that while Nb and Fe stay close to the surface, Ag intercalates into the van der Waals gap of the topological insulator. At low temperatures, for all three evaporants, a reversal in the effect on the BSTS band alignment is seen. This means  a switch from donor-like for $\leq$0.1ML ($\leq$0.2ML for Fe), to acceptor-like for thicker deposited overlayers. This has not been reported in analogous studies on either Bi$_2$Se$_3$ or Bi$_2$Te$_3$, and could help explain some seemingly contradictory results in the literature, such as the different doping behavior that has been reported for studies of only very low \cite{Honolka2012I} or higher coverages  \cite{Scholz2012T} of Fe on Bi$_2$Se$_3$. Our identification of a reversal point around 0.1 ML between low-T n- and p-type behavior for different evaporants on BSTS lies exactly in the coverage gap between these two studies on Fe/Bi$_2$Se$_3$. In analogy with the effect of Au overlayers on a band-bent oxide surface  \cite{Zhang2011E}, cluster formation on the surface at low T could be the driving force for the observed reversal in band energy shifts for coverages $>$0.1 ML.\\
\indent
In the particular case of silver, formation of Ag$_{\text{Te,Se}}$ defects, possibly aided by oxygen-related effects, are additional drivers for p-type behavior. That this reversal in doping character is not observed when Nb or Fe deposition is carried out at room temperature is consistent with the preference at higher temperatures for interstitial and sub-surface sites over Õon-topÕ cluster formation.\\
\indent
Core level data taken at room temperature shows that at high coverage all three species of adatoms form substitutional defects in the Te/Se layer. These substitutions are not observed to have a significant effect on the electronic structure of the surface. The picture is different at 38K: where it becomes clear that the adatoms are essentially frozen into place on the surface, unable to move into the most energetically favorable position, meaning that Nb and Fe do not form any substitutional defects even at high coverage. In contrast, Ag is able to move and form substitutional defects also at 38K, but is not mobile enough at this temperature to intercalate into the van der Waals gap as it does at room temperature.\\
\indent
From these observations we show how the mobility and the density of the metal adatoms play a very important role in determining the influence of interface creation on the electronic properties of the bulk insulating TI BSTS. N-type or p-type behavior can be achieved at will. For the most commonly used interface formation method involving evaporation onto a substrate held at room temperature, we show n-type doping to be the result. However, the cryogenic experiments also yield a clear conclusion that on increasing deposition at low temperature, the doping character of Ag, Nb and Fe can be reversed. This low-T, p-type behavior can also go far enough to depopulate the topologically trivial conduction band in the near-surface region. The total energy range available in which the Dirac point energy can be tuned using a combination of metal coverage and sample temperature is shown to be between 150-350 meV below the Fermi level.

\section*{Acknowledgements}    
This work is part of the research program of the Foundation for Fundamental Research on Matter (FOM), which is part of the Netherlands Organization for Scientific Research (NWO). E.v.H. acknowledges support from the NWO VENI program.\\ \indent The research leading to these results has received funding from the European Community's Seventh Framework Program (FP7/2007-2013) under grant agreement no. 312284.

\renewcommand\thefigure{A1}
\setcounter{figure}{0}

\begin{figure*}
	\begin{center}
		\centering
		\includegraphics[width = 15 cm]{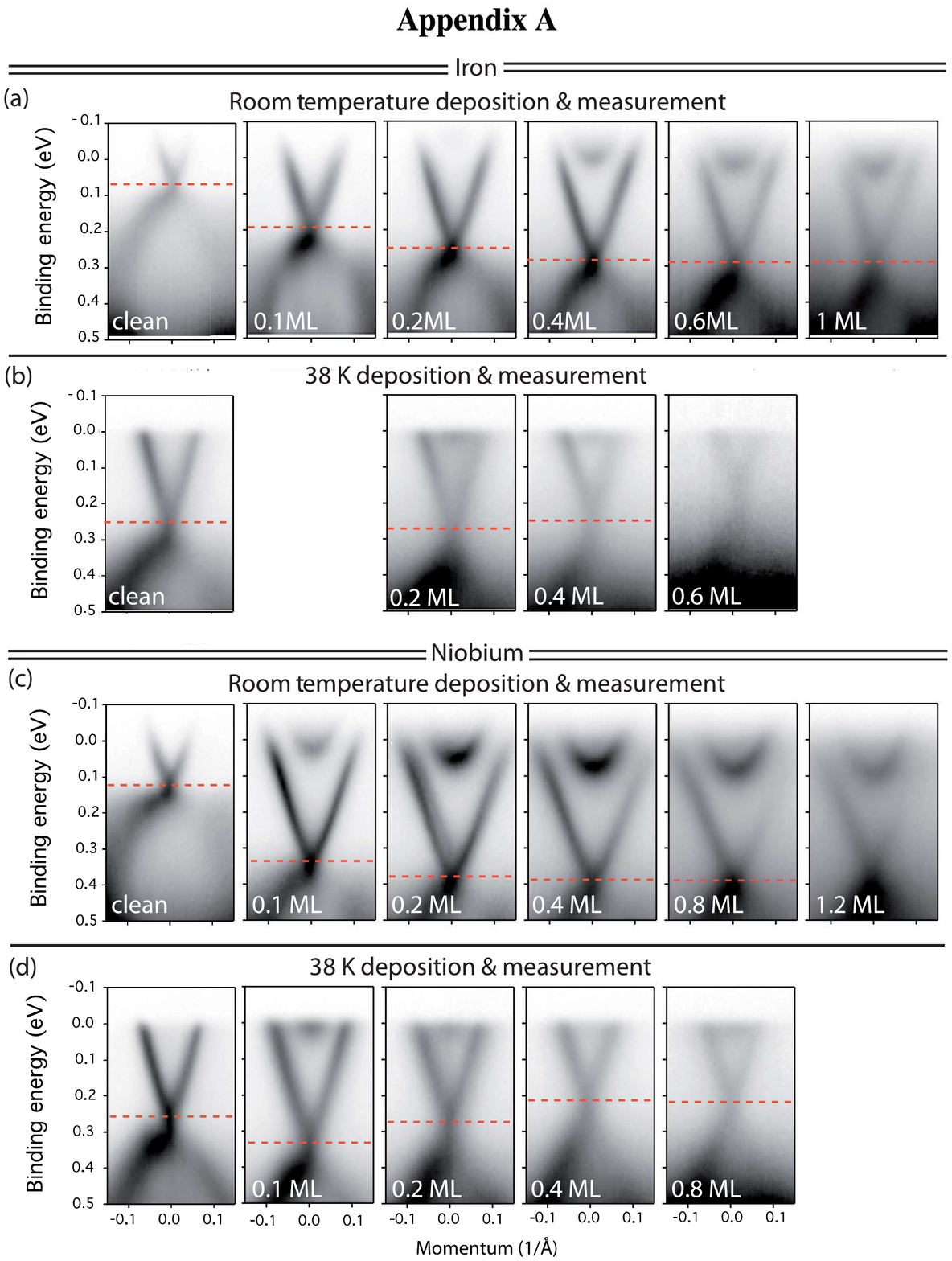}\\
	\end{center}
	\caption{
		Near-E$_F$ electronic structure of Bi$_{1.5}$Sb$_{0.5}$Te$_{1.7}$Se$_{1.3}$ with different amounts of Nb and Fe deposited on the surface at either room temperature of 38K. (a) Deposition of iron at room temperature for the coverage of 0, 0.1, 0.2, 0.4, 0.6 and 1.0 ML. (b) Deposition of iron at 38K for the coverage of 0, 0.2, 0.4 and 0.6. (c) Deposition of niobium at room temperature for the coverage of 0, 0.1, 0.2, 0.4, 0.8 and 1.2 ML. (d) Deposition of niobium at 38K for the coverage of 0, 0.1, 0.2, 0.4 and 0.8 ML. All measurements were performed using a photon energy of 27 eV. 
	}
\end{figure*}

\end{document}